\documentclass[12pt]{article}
\usepackage{amssymb,amsmath,epsfig}

\begin{document}

\title{\bf Cosmological Evolution Via Interacting/Non-Interacting Holographic Dark Energy Model
for Curved FLRW Space-time in Rastall Gravity}
\author{Rabia Saleem\thanks{rabiasaleem@cuilahore.edu.pk} and Shahnila
\thanks{shahnilapervaiz183@gmail.com} \\
Department of Mathematics, COMSATS University Islamabad,\\
Lahore Campus, Lahore-54590, Pakistan.}

\date{}
\maketitle

\begin{abstract}

In this work, we explore the phenomenon of cosmic evolution using curved FLRW space-time
bounded by apparent horizon with a specific holographic cut-off. To this end, we use the
framework of Rastall gravity (RG) and universe is assumed to be consists of interacting/
non-interacting dark energy (DE) and dark matter (DM). In both scenarios, we evaluate exact
solutions of the dynamical equations and constraint the holographic parameter $c^{2}(z)$
assuming a slowly varying function of red-shift during non-interacting model. For interacting
model, we consider $c^{2}$ as a constant function of red-shift. Moreover, we analyze nature of
the obtained results via deceleration parameter $(q)$, statefinder pair $(j,s)$ and $Om(z)$-diagnostic
by constraining the involved model parameters using latest observational data. The graphical analysis
showed that interacting model is very close to $\Lambda$ cold DM ($\Lambda$CDM) model as compared
to non-interacting case. We conclude that this holographic proposal is enough to describe the cosmic
evolution at an accelerating rate.
\end{abstract}


\section{Introduction}

Cosmic evolution can be defined as the study of evolution, the vast number of the experimental and productive
changes that have been collected during all the time and across all space from big bang to humankind. With the
passage of time, cosmic evolution involves in the theory of general relativity (GR). There are many proposed
models that follow accelerated cosmic phase but some observations can not be explained by these models such as,
the rotation curves of galaxies, theoretical problems of the cosmological constant $(\Lambda)$ as well as the
possible values of local anisotropy in the evolution of the universe. To tackle these problems, we need modified
theories of gravity that provide a foundation for current understanding of the physical phenomena happening in
the universe. For detail study of modified theories of gravity, see reference \cite{1}.

In literature, it can be seen that accelerating expansion is due to some tropical cosmic component known
as DE. It was discovered in 1998 by two international teams (High-Z supernova search team and Type 1a supernova
(\textit{SN1a}) team) in which American Astronomers Riess and Perlmutter and Australian Astronomer Schmidt included.
Dark energy is the name given to a mysterious component that cause the rate of the expansion of universe to accelerate
over time rather than to slow down. It is an unknown form of energy that affects the universe at largest scales, acting
opposite to gravity. In literature, astrophysical observations of \textit{SN1a} \cite{2}, large scale structure
(\textit{LSS}) \cite{3}, cosmic microwave background (\textit{CMB}) \cite{4} and planck measurements \cite{5} indicate
that our universe is in accelerating phase. There are many proposed DE models that can explain the cosmic nature, such
as dynamical DE models (including \textit{holographic DE (HDE)} model, \textit{Ricci DE (RDE)} model,
\textit{Tsallis HDE} model, small positive $\Lambda$ etc.) \cite{6} and many types of scalar field models
like \textit{phantom}, \textit{quintessence}, \textit{k-essence} etc. \cite{7}-\cite{13}.

In 1970, Weinberg introduced an equation of state (EoS) which is the ratio of total energy density $(\rho)$ to the
total pressure $(P)$ of the fluid i.e., $\omega=\frac{P}{\rho}$, where $\omega$ be the dimensionless EoS parameter.
This parameter characterizes different phases of the universe such as cosmological constant era for which $\omega$
is always equal to -1. If DE have not shown as $\Lambda$ then the next phases are the \textit{phantom} phase in
which $\omega<-1$ \cite{10}, \textit{quintessence} for $\omega>-1$ and \textit{quintom} (that can cross the
\textit{phantom} divide line) \cite{11}. Equation of state parameter for DE varies along red-shift parameter $(z)$.
According to Planck collaboration 2020 results, EoS parameter for DE is measured to be $\omega=-1.03\pm0.03$,
which is consistent with $\Lambda$.

In this paper, we approach our problem for DE using holographic principle which can figure out current cosmic acceleration.
When the region is finite then this principle shows that the degrees of freedom of the physical system should finite \cite{12,13}.
Cohen et al. \cite{28} proposed that there is a relation between ultraviolet (UV) cut-off and infrared (IR) cut-off in the quantum
field theory. This relation occurs during the evolution of a black hole (BH), namely, if $\rho_{\Lambda}$ which is the quantum zero
point energy density generated by UV cut-off, then the total energy having size $L$ in the region must be less then the mass
of the BH of the same size, so $L^{3}\rho_{\Lambda}\leqslant LM_{p}^{2}$. At largest value of $L$, it will saturate an inequality, i.e., $\rho_{\Lambda}=3c^{2}M_{p}^{2}L^{-2}$ and a factor $c^{2}$ appeared. We deal $3c^{2}$ as a numerical constant.
Li \cite{15} proposed a HDE model, which gave the idea that UV cut-off is related to the IR cut-off and
proved that HDE scheme is feasible if we adjust the IR cut-off by the event horizon. This is not only a feasible model, but it
also makes a physical prediction about the EoS of DE, which is applicable to future experiments. Sharif et al. \cite{16} studied
new HDE model in Brans-Dicke theory for interacting and non-interacting cases and proved that their results are compatible with
the corresponding isotropic universe model. In literature, Mamon et al. \cite{14} worked with HDE model in the framework of
cosmology and considered an interaction between HDE and a pressure-less DM. In this scenario, they proved that HDE model is
entered into accelerating era at a specific value $z=0.83$.

In previous studies, a large number of interaction terms have been defined some of them are in the following form:
$Q\varpropto H\rho_{1}$, ~$Q\varpropto H\rho_{2}$ and $Q\varpropto H(\rho_{1}+\rho_{2})$, here $\rho_{1}$ stand for
DE density and $\rho_{2}$ is the DM density \cite{22}. We decided to work with the first interaction term
$Q=b^{2}H\rho_{1}=\Gamma\rho_{1}$, where $b^{2}$ is the free coupling parameter. In this context, Arevalo et al. \cite{23}
worked with Ricci cut-off taking flat FLRW space-time and involved RDE-DM interaction and checked the behavior of universe
by computing some essential cosmological parameters. Saleem and Imtiaz \cite{24} studied interacting dynamical RDE model
for curved FLRW space-time. To evaluate the exact solution of the model, they used Chevallier-Polarsky-Linder type
parametrization for coincidence parameter $r(z)$. In their work, they discussed about future singularity and formulated
some cosmological parameters and compared their results with $\Lambda$CDM model. For detail study of this topic
see reference \cite{25}-\cite{27}.

Moreover, we will work on cosmological parameters involving deceleration parameter $q$, statefinder pair $(j,s)$ and
$Om(z)$-diagnostic. It is well-known that universe is expanding but the rate at which it is expanding is also important
to know. This expansion rate is due to the Hubble parameter $(H)$. Basically the parameter $q$ is a way of evaluating this.
This parameter appears in the Taylor series expansion of the scale factor $a(t)$, if we discuss $a(t)$ at the present time
the general expression becomes $a(t)= a(t_{0})+ \dot{a}(t_{0})(t-t_{0})+\frac{1}{2}\ddot{a}(t_{0})(t-t_{0})^{2}+\cdots$,
if we split up this term by $a(t_{0})$ then the coefficient of the term $(t-t_{0})$ becomes $H$ and the expression turn out to be
$a(t)a(t_{0})^{-1}=1+H_{0}(t-t_{0})-\frac{q_{0}}{2}H_{0}^{2}(t-t_{0})^{2}+\cdots$. Here $q_{0}=-a(t)\ddot{a}(t_{0})(\dot{a}^{2}(t_{0}))^{-1}= -1-\dot{H}(H^{2})^{-1}$. The parameter $q$ is utilized to discover the nature of cosmos whether it is accelerating, decelerating or static.
Cruz and Lepe \cite{30} found $q$ and proved that it must fell in $q(0)=[-1.01,0.07]$ and important argument is that this interval included the $\Lambda$CDM model. The parameters $H$ and $q$ can effectively describe the behavior of cosmos but are not the best indicators to
discriminate among various DE models.

Sahni et al. \cite{31} introduced a dimensionless pair $(j,s)$ which can discriminate among many DE models and has ability to check
how a model is far from the $\Lambda$CDM model. Jawad et al. \cite{32} found $(j,s)$ and proved that some results follow DE eras, like \textit{phantom} and \textit{quintessence} and some results approach the $\Lambda$CDM limit, $(j,s)=(1,0)$. Jerk parameter $(j)$ is
constructed using third derivative of $a(t)$ with respect to time while snap parameter $(s)$ contains its fourth derivative.
$Om(z)$ is a geometrical diagnostic that develop a relation between $H$ and $z$ and can be described for zero, positive and negative
curvature for $\Lambda$CDM, \textit{phantom} and \textit{quintessence}, respectively. Shahalam et al. \cite{33} applied $Om(z)$-diagnostic
on scalar field models and showed that their results led to cosmic acceleration.

Rastall gravity was developed in 1972 and now a days it becomes an eminent theory of gravity because it is presented as a
modified theory of gravity with non-conserved stress energy tensor and an unusual non-minimal coupling between geometry and
matter. In GR, the conservation law that showed a divergence free energy momentum tensor has been approved i.e.,
$\nabla^{\nu}T_{\mu\nu}=0$, where $\nabla^{\nu}$ is the covariant derivative. On contrary, Rastall adopted a novel and
different conservation law that hypothesized the GR ideas. In this theory, the energy momentum tensor does not a conserved
quantity. Rastall theorem is defined as $\nabla^{\mu}T_{\mu\nu}=(\frac{\kappa}{16\pi})\nabla_{\nu}R$, where $\kappa$ is
the coupling constant and $R$ be the Ricci scalar. The second Bianchi identity in this theory remains the same i.e.,
$ \nabla^{\nu}G_{\nu\mu}=0$, where $G_{\mu\nu}=R_{\mu\nu}-\frac{1}{2}Rg_{\mu\nu}$ be the Einstein tensor. An important
fact of RG is that the modifications have been done in the matter part of the theory only keeping the mathematical part
invariant \cite{17}-\cite{19}. Das et al. \cite{20} presented some cosmological consequences in the structure of modified
RG and proved that RG is equivalent to the Einstein gravity. Ghaffari et al. \cite{21} presented HDE in RG, considering
vacuum energy that plays the role of DE. They adopted the phenomenon where the sum of this energy and Rastall term is
being responsible for the current accelerating universe.

In this work, we consider two dark components as HDE and DM for cosmic fluid. We study two different cases: first is
the non-interacting HDE-DM model which means that these two components are self-conserved. In second case, we work with
the interacting model introducing a specific interaction term $Q$ between these two dark components. We check physical
viability of both of the models making complete physical analysis. The following work is arranged as under: In section
\textbf{2}, we develop the dynamical equations including field equations and corresponding conservation equations for
curved FLRW space-time for interacting and non-interacting HDE-DM within RG. In section \textbf{3}, the exact solutions
for normalized Hubble parameter $H_{n}^{2}(z)$, coincidence parameter $r(z)$ and $\rho_2$ are evaluated for non-interacting
model and graphically analyze the behavior of these parameters. We constraint slowly varying function $c^{2}(z)$ and check
that this range describe the cosmic evolution. We study some cosmological consequences like $\omega$ and $q$ in term of
red-shift. In section \textbf{5}, we re-evaluate the considered model for interacting case and plot their trajectories.
For both cases, we find $q$, $(j,s)$ and $Om(z)$-diagnostic and analyze their behavior graphically in section \textbf{6}.
The last section conclude the results.

\section{Basic Dynamical Equations in Rastall Gravity}

In this section, we use framework of RG to explain the dynamics of two interacting and non-interacting components of the
fluid that are HDE and DM. The geometry of curved FLRW space-time has been under consideration which is given by
\begin{equation}\nonumber
ds^2=-dt^2+a(t)^2\bigg(\frac{dr^2}{1-Kr^2}+r^2d\Omega^2\bigg),
\end{equation}
where $d\Omega^{2}=d\theta^{2}+\sin^{2}\theta d\phi^{2}$ and the curvature parameter $K=0,~\pm1$ for flat, closed and
open universe, respectively. The Rastall field equations are developed as under to find the Friedmann constraint for a
perfect fluid
\begin{eqnarray}\nonumber
3(1-4\kappa\lambda)H^2-6\kappa\lambda\dot{H}+3(1-2\kappa\lambda)\frac{K}{a^2}&=&\kappa\rho,\\\nonumber 3(1-4\kappa\lambda)H^2+2(1-3\kappa\lambda)\dot{H}+(1-6\kappa\lambda)\frac{K}{a^2}&=&-\kappa P,
\end{eqnarray}
in above two equations, dot represents derivative with respect to time, $\lambda$ is an additional parameter, $P=P_1+P_2$ and
$\rho=\rho_1+\rho_2$ are total pressure and total energy density of the fluid, respectively. Whereas the subscript 1 and 2
associated to DM and HDE, respectively.

Friedmann constraint can be evaluated as follows using standard EoS, $P=\omega\rho$
\begin{equation}\label{1}
3H^2(1-4\kappa\lambda)(1-\Omega_{K})=\rho_{1}\kappa(1+r)(1-3\kappa\lambda(1+\omega)),
\end{equation}
in this equation, $r$ is the coincidence parameter which is define as $r=\frac{\rho_{2}}{\rho_{1}}$ and
\begin{equation}\nonumber
\Omega_{K}(z)=-\frac{K}{a_{0}^2H_{0}^2}(1+z)^2=\Omega_{K}(0)(1+z)^2,
\end{equation}
here $a_{0}$ is the scale factor and $H_{0}$ is the Hubble constant at the present time $(z=0)$. The following
relation is used to evaluate Eq. (\ref{1})
\begin{equation}\nonumber
(1+z)= a_{0}a^{-1}.
\end{equation}
Next, we will develop the main formalism of non-interacting scenario.

\section{Dynamics of Non-interacting Model}

Conservations equations for the DE and DM components of the fluid are calculated in the form of $\rho_1$,
respectively as under
\begin{eqnarray}\label{2}
\rho_{1}^\prime\bigg(\frac{1-3\kappa\lambda-3\kappa\lambda\omega_{1}}{1-4\kappa\lambda}\bigg)
-3\bigg(\frac{1+\omega_{1}}{1+z}\bigg)\rho_{1}&=&0,\\\label{3} (r\rho_{1})^\prime\bigg(\frac{1-3\kappa\lambda-3\kappa\lambda\omega_{1}}{1-4\kappa\lambda}\bigg)
-3\bigg(\frac{1+\omega_{2}}{1+z}\bigg)(r\rho_{1})&=&0,
\end{eqnarray}
here prime denotes derivative with respect to red-shift, and $\omega_{1},~\omega_{2}$ are the EoS parameters
for DE and DM, respectively.

\subsection{Holographic Cut-off as Apparent Horizon and $c^2(z)$ Function}

The HDE model is one of the remarkable attempts to examine the nature of DE in the framework of quantum
gravity through the holographic principle. Based on this principle, we deal with $\rho\sim L^{-2}$,
where $L$ denotes the size of the universe. For curved space-time, the HDE has the energy density
given as under
\begin{equation}\label{4}
\rho_{1}=3(\beta_{1}-\beta_{2}\Omega_{K})H^2,
\end{equation}
here $\beta_{1}$ and $\beta_{2}$ are constant parameters. The radius of the apparent horizon is defined as
\begin{equation}\nonumber
r_{ah}=\frac{1}{\sqrt{(1-\Omega_{K})H^2}}.
\end{equation}
Holographic principle is suggested that particle entropy within cosmological apparent horizon should not be
greater than gravitational entropy, which is related to the apparent horizon \cite{34}. Energy density given
in Eq. (\ref{4}) can be expressed in terms of red-shift as follows
\begin{equation}\label{5}
\rho_{1}(z)=3(\beta_{1}-\beta_{2}\Omega_{K}(0)(1+z)^2)H^2(z).
\end{equation}
Observe that, when $z\rightarrow -1$ then $\rho_{1}\rightarrow 3\beta_{1}H^2$. This $H^2$ is different from
the proposal of Li. Li has rejected the identification of IR cut-off with Hubble's radius \cite{29}. Taking
$L=\frac{1}{H}$, implying that holographic bound is saturated, the energy density becomes $\rho_{1}=3c^2M_{P}^{2}H^2$,
where $H^2$ is the only constant given by $3c^2$. Usually HDE density contains a constant term $c^2$ which lie in
the interval $0<c^2<1$. We consider HDE model with $c^2(z)$ parameter that provide a cosmological constant, cosmic
expansion or the eternal expansion of the universe.

We can rewrite the expression of $\rho_{1}(z)$ given in Eq.(\ref{5}) in terms of $c^2(z)$ as
\begin{equation}\label{6}
\rho_{1}(z)=3c^2(z)H^2(z),
\end{equation}
where $c^2(z)$ is assumed to be
\begin{equation}\label{7}
c^2(z)=\beta_{1}-\beta_{2}\Omega_{K}(0)(1+z)^2.
\end{equation}
At present time $z=0$, we see that the function $c^2(0)=\beta_{1}$ is a pure constant, this holds for flat universe
when $\Omega_{K}(0)=0$. Using value of $\rho_{1}(z)$ given in Eq.(\ref{6}) in Eq.(\ref{1}), we get the following
expression
\begin{equation}\label{8}
c^2(z)=\bigg(\frac{1-4\kappa\lambda}{\kappa(1-3\kappa\lambda(1+\omega))}\bigg)\bigg(\frac{1-\Omega_{K}(z)}{1+r(z)}\bigg).
\end{equation}
This equation will help out to constraint $c^{2}(z)$ at the present time. Via Eq.(\ref{5}) for $\rho_{1}(z)$, we can
evaluate $\rho_{2}(z)$ in the following form
\begin{equation}\label{8a}
\rho_{2}(z)=((1+z)^{3})\Omega_{2}(0))^{\frac{1-4\kappa\lambda}{1-3\kappa\lambda}}.
\end{equation}
Equations (\ref{5}) and (\ref{8a}) lead us to develop the expression of $r(z)$ at present time as
\begin{equation}\nonumber
r(0)=\frac{(\Omega_{2}(0))^{\frac{1-4\kappa\lambda}{1-3\kappa\lambda}}}{3H_{0}^{2}(\beta_{1}-\beta_{2}\Omega_{K}(0))}.
\end{equation}
Substituting the above value back in Eq.(\ref{8}), we get $c^{2}(z)$ at $z=0$
\begin{equation}\label{8aa}
\frac{3}{2}c^{2}(0)=\bigg(\frac{1-4\kappa\lambda}{\kappa(1-3\kappa\lambda(1+\omega_{0}))}\bigg)(1-\Omega_{K}(0))
-\frac{(\Omega_{2}(0))^{\frac{1-4\kappa\lambda}{1-3\kappa\lambda}}}{3H_{0}^{2}}+\beta_{2}\Omega_{K}(0).
\end{equation}

\begin{table}[htb]
\centering
\begin{tabular}{|c | c | c | c | c | c | c|}
\hline
\multicolumn{4}{|c}{$\lambda=10^{-1}$} & \multicolumn{3}{|c|}{$\lambda=10^{-2}$}\\
  \hline
  $\Omega_{K}(0)$&0.005 & -0.005 & 0 & 0.005 & -0.005 & 0 \\
  \hline
  $c^{2}(0)$&0.64421 & 0.64407 & 0.64418 & 0.63955 & 0.63927 & 0.63937 \\
  \hline
\end{tabular}
\caption{Values of $c^{2}(0)$ for open, closed and flat universe.}
\end{table}
\begin{table}[htb]
\centering
\begin{tabular}{|c | c | c | c | c | c | c|}
\hline
\multicolumn{4}{|c}{$\lambda=10^{-3}$} & \multicolumn{3}{|c|}{$\lambda=10^{-4}$}\\  \hline
  $\Omega_{K}(0)$&0.005 & -0.005 & 0 & 0.005 & -0.005 & 0 \\
  \hline
  $c^{2}(0)$&0.66394 & 0.66391 & 0.66392 & 0.66638 & 0.66637 & 0.66637 \\
  \hline
\end{tabular}
\caption{Values of $c^{2}(0)$ for open, closed and flat universe.}
\end{table}
According to Planck collaboration 2020 results \cite{5}, $H_{0}=67.4\pm0.5$ and $\omega_{0}=-1.03\pm0.03$. We pick the
value of $\Omega_{2}(0)=0.3089\pm0.0062$ from \cite{30} and $\Omega_{K}(0)=\pm0.005^{K=-1}_{K=1}$ for open and closed
universe, respectively and $\Omega_{K}(0)=0$ for flat universe from \cite{24}. Now varying the theory parameter $\lambda$,
we find different corresponding values of $c^{2}(0)$ from Eq.(\ref{8aa}), which further leads us to find a feasible interval
for $c^{2}(0)$. The values of parameter $c^{2}(0)$ are calculated in Table \textbf{1} and \textbf{2}.
From these tables, we can see that as $\lambda<<1$, $c^{2}(0)$ lies between $[0.63,0.66]$. For large values of $\lambda$,
negative values for $c^{2}(0)$ appear that is not a physical case. Now we write Eq.(\ref{7}) in just one free parameter,
i.e., $\beta_{1}$ as
\begin{equation}\label{9}
c^2(z)=\beta_{1}+(c^2(0)-\beta_{1})(1+z)^2,
\end{equation}
putting this value of $c^2(z)$ in Eq.(\ref{6}), we get $\rho_{1}(z)$ in the following form
\begin{equation}\label{10}
\rho_{1}(z)=3(\beta_{1}+(c^2(0)-\beta_{1})(1+z)^2)H^2(z).
\end{equation}

Now, we will develop a relationship between $c^2(z)$ and constant parameter $\beta_{1}$, which appear
in the holographic cut-off. In general, $c^2$ can be considered as a slowly varying function of time,
the term $\frac{\dot{c^2}}{c^2}$ has Hubble expansion rate $H$ as an upper bound \cite{35}
\begin{equation}\label{11}
\frac{\dot{c^{2}}(t)}{c^2(t)}\lesssim H,
\end{equation}
using Wetterich parametrization in which holographic parameter $c^2$ can be taken in terms
of $z$, the above equation turns out to be
\begin{equation}\nonumber
(1+z)\frac{1}{c^2(z)}\frac{dc^2(z)}{dz}\gtrsim -1.
\end{equation}
Substituting Eq.(\ref{9}) in Eq.(\ref{10}), we obtain the following relation
\begin{equation}\label{12}
-3(c^2(0)-\beta_{1})(1+z)^2\lesssim \beta_{1},
\end{equation}
which further reduced to following form at the present time
\begin{equation}\label{14}
\beta_{1}\lesssim \frac{3}{2}c^2(0).
\end{equation}
For all $z$, we have condition listed under
\begin{equation}\nonumber
c^2(z)\gtrsim \frac{2\beta_{1}}{3}\quad\Rightarrow\quad\frac{2}{3}\beta_{1}\lesssim c^2(z)<1.
\end{equation}
From Eq. (\ref{12}), we find the following expression
\begin{equation}\label{15}
1+z\lesssim \sqrt{\frac{\beta_{1}}{3(\beta_{1}-c^2(0))}},
\end{equation}
which suggests that $\beta_{1}$ must be greater than $c^2(0)$ to obtain the real valued function.
Hence, we can find the range of $\beta_{1}$ given as
\begin{equation}\label{16}
c^2<\beta_{1}\lesssim\frac{3}{2}c^2(0).
\end{equation}
Taking $z=-1$ in Eq.(\ref{15}), slowly varying condition results $\beta_{1}\gtrsim 0$, which
further constrain $\beta_{1}$ as $0\lesssim\beta_{1}\lesssim\frac{3}{2}c^2(0)$. If we use
value of $\beta_1$ from Eq.(\ref{14}) in Eq.(\ref{15}), we get $z\lesssim0$, which shows that
holographic cut-off for DE density is enough to describe cosmic evolution.

In coming subsections, we will evaluate exact solutions of the dynamical equations including normalized
Hubble parameter, coincidence parameter and EoS parameter $\omega_1$ for non-interacting case.

\subsection{Normalized Hubble Parameter for Non-interacting Model}

Using first Friedmann equation (\ref{1}), we derive $H_{n}^{2}(z)$ in the following form
\begin{equation}\nonumber
H_{n}^{2}(z)=\frac{\kappa(1-3\kappa\lambda(1+\omega))((1+z)^3\Omega_{2}(0))^\frac{1-4\kappa\lambda}
{1-3\kappa\lambda}}{3H_{0}^2(1-4\kappa\lambda)
(1-(1+z)^2\Omega_{K}(0))(1-\frac{\kappa(1-3\kappa\lambda(1+\omega))c^2(z)}{(1-4\kappa\lambda)
(1-(1+z)^2\Omega_{K}(0))})},
\end{equation}
after inserting the expression of $c^2(z)$, the above equation turn out to be
\begin{eqnarray}\nonumber
H_{n}^{2}(z)&=&\bigg(1-\frac{\kappa(1-3\kappa\lambda(1+\omega))(\beta_{1}-(1+z)^2(\beta_{1}-c^2(0)))}{(1-4\kappa\lambda)
(1-(1+z)^2\Omega_{K}(0))}\bigg)^{-1}\\\label{17}&\times&\bigg(\frac{\kappa(1-3\kappa\lambda(1+\omega))
((1+z)^3\Omega_{2}(0))^\frac{1-4\kappa\lambda}
{1-3\kappa\lambda}}{3H_{0}^2(1-4\kappa\lambda)(1-(1+z)^2\Omega_{K}(0))}\bigg),
\end{eqnarray}
here $H_{n}^{2}(z)=\frac{H(z)}{H(0)}$.
\begin{figure}\centering
\epsfig{file=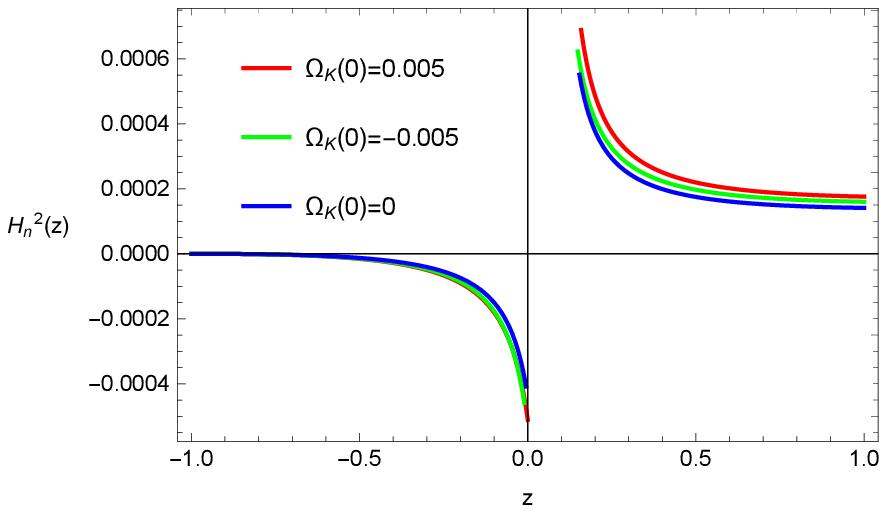, width=0.50\linewidth}\epsfig{file=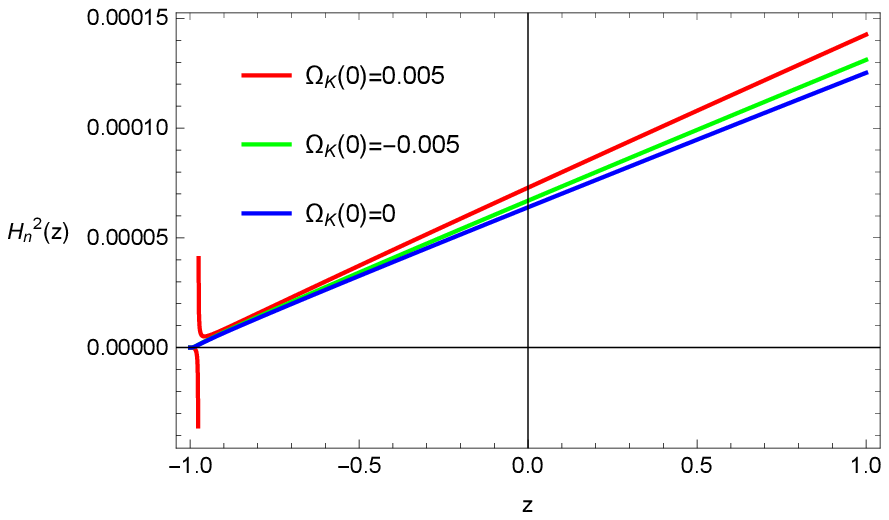, width=0.50\linewidth}
\caption{Graph of $H_{n}^{2}(z)$ versus $z$. Left plot for $\lambda=10^{-1}$ and right plot for $\lambda=10^{-2}$.
Values of the remaining parameters are taken from Table \textbf{1}. \label{f1}}
\end{figure}
\begin{figure}\centering
\epsfig{file=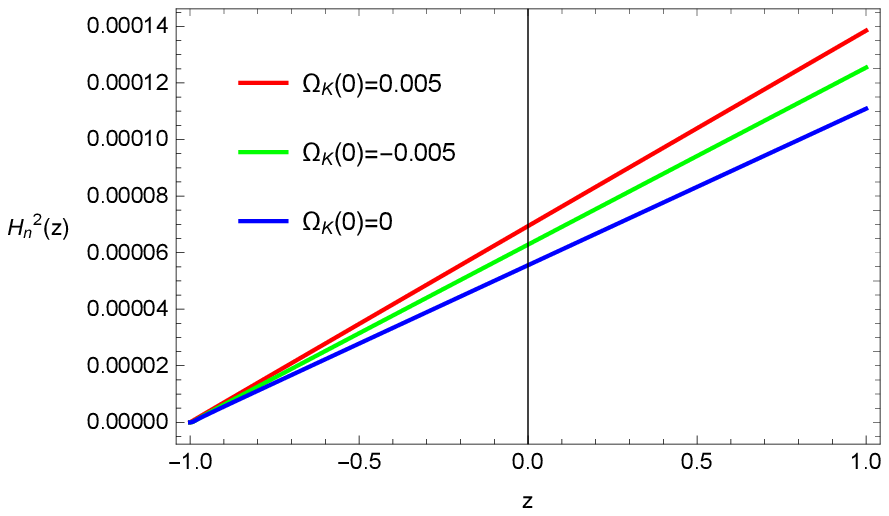, width=0.50\linewidth}\epsfig{file=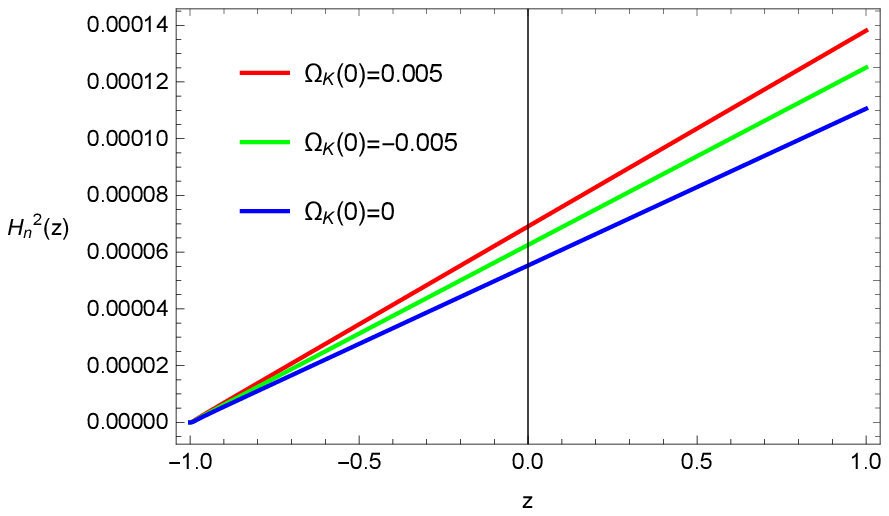, width=0.50\linewidth}
\caption{Graph of $H_{n}^{2}(z)$ versus $z$. Left plot for $\lambda=10^{-3}$ and right plot for $\lambda=10^{-4}$.
Values of the remaining parameters are taken from Table \textbf{2}. \label{f2}}
\end{figure}
It is observed that the expression of $H_{n}^{2}(z)$ depend upon eight parameters namely $\Omega_{K}(0),~\Omega_{2}(0),~\kappa,~\omega,~\beta_{1},~H_{0},~\lambda$ and $c^2(0)$. Using values of these parameters
as specified in the Tables \textbf{1} and \textbf{2}, we analyze the obtained result graphically as shown in Figs. \textbf{1}
and \textbf{2}. Trajectories for $\lambda=10^{-1}$ showing that as model approaches to far future, $H_{n}^{2}(z)$ increases,
this behavior differs from $\Lambda$CDM model in which $H$ has a bounded value. While in the past, $H_{n}^{2}(z)$ has positive
decreasing range for all the three types of geometries. For $\lambda=10^{-2}$ (right plot of Fig. \textbf{1}), it shows an
increasing behavior as $z$ move from past to future epoch. On the similar manner, Fig. \textbf{2} (for $\lambda=10^{-3}$
(left panel) and for $\lambda=10^{-4}$ (right panel)) is exhibiting an increasing behavior from past to future epoch for closed,
flat and open universe model. Hence, we can conclude that as $\lambda$ should be small, i.e., $0\leq\lambda<1$ to show expansion
of the universe otherwise, results are not feasible. It is verified that for $\lambda=0$, $H_{n}^{2}(z)$ obey physical behavior
for closed, flat and open universe.

\subsection{Coincidence Parameter for Non-interacting Model}

To evaluate the value of coincidence parameter, we consider the results of $\rho_{1}(z)$ and $\rho_{2}(z)$, and $r(z)$ has
the following form
\begin{equation}\label{18}
r(z)=\frac{\rho_{2}(z)}{\rho_{1}(z)}=\frac{((1+z)^3\Omega_{2}(0))^\frac{1-4\kappa\lambda}{1-3\kappa\lambda}}
{3c^2(z)H^{2}_{n}(z)H_{0}^2}.
\end{equation}
Substituting the value of $H_{n}^{2}(z)$ and $c^{2}(z)$, Eq.(\ref{18}) becomes
\begin{eqnarray}\nonumber
r(z)&=&\bigg(\frac{1-(\frac{\kappa-3\kappa^2\lambda(1+\omega)}{(1-4\kappa\lambda)(1-\Omega_{k}(0)(1+z)^2)})
(\beta_{1}-(\beta_{1}-c^{2}(0))(1+z)^{2})} {\beta_{1}-(\beta_{1}-c^{2}(0))(1+z)^{2}}\bigg)\\\label{19}&\times& \bigg(\frac{(1-4\kappa\lambda)(1-\Omega_{K}(0)(1+z)^2)}{\kappa-3\kappa^2\lambda(1+\omega)}\bigg).
\end{eqnarray}
Putting specific values of all the involved model parameters from Tables \textbf{1} and \textbf{2}
in the above expression, we observed that at present time $r(0)<1$, which generate the following
inequality to constraint $c^2(0)$
\begin{equation}\label{20}
\frac{1}{2}<c^2(0)<1,
\end{equation}
hence this condition remains consistent with the closed interval $[0.63,0.66]$.

\begin{figure}\centering
\epsfig{file=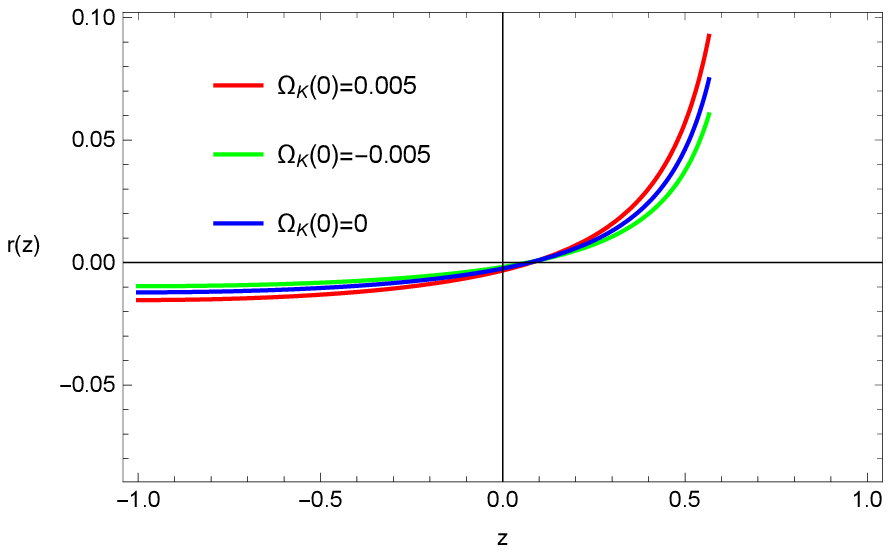, width=0.50\linewidth}\epsfig{file=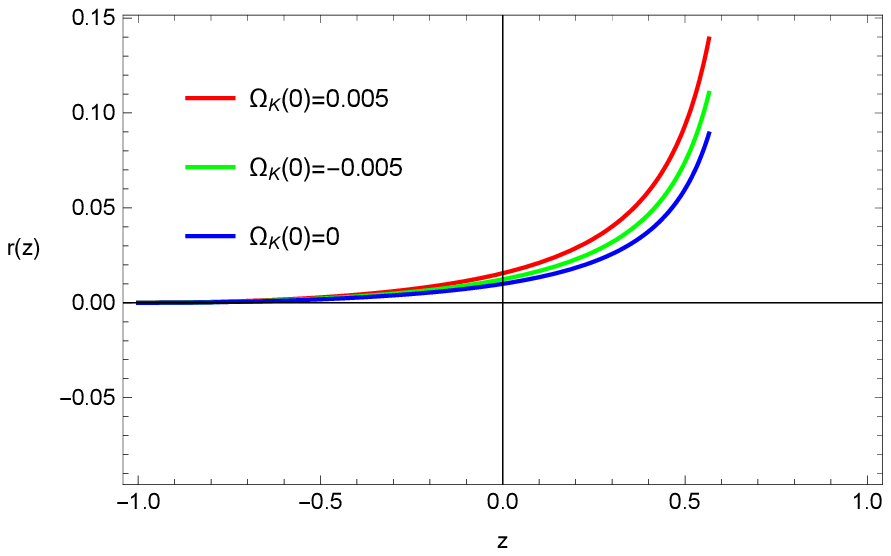, width=0.50\linewidth}
\caption{Graph of $r(z)$ versus $z$ for $\lambda=10^{-1}$ and $\lambda=10^{-3}$. \label{f3}}
\end{figure}

The behavior of $r(z)$ using specific values of involved parameters has been shown in Fig. \textbf{\ref{f3}}.
It is well-known that at present time, the density ratio is of the order of unity or we can also say that both
densities have the same order of magnitude i.e., $\frac{\rho_{2}}{\rho_{1}}\thicksim O(1)$. This phenomenon is
known as coincidence problem and it also indicates that we attain a special period of cosmic evolution. The coincidence
parameter is not much sensitive for $\lambda$ as it behaves the same for both small as well as large values of $\lambda$.
The parameter $r(z)$ decreases as universe evolves from past to future era for closed, flat and open universe. It can
be seen from Eq.(\ref{19}) that $r(z)$ is independent of $\Omega_{2}(0)$ parameter, which is the clear evidence that
it represents DE dominant era and it is also known as feature of cosmological coincidence problem \cite{36}. For
$\lambda=0$, it shows the same physical behavior for closed, flat and open universe.

\subsection{Equation of State Parameter $\omega_{1}(z)$}

In this section, we will evaluate $\omega_{1}(z)$ for DE component and analyze it graphically. We use the
conservation equation (\ref{2}) for DE to find the expression of this parameter as
\begin{equation}\label{21}
1+\omega_{1}(z)=\frac{1}{3(1-4\kappa\lambda)\frac{\rho_{1}}{\rho^{\prime}_{1}}+3\kappa\lambda}.
\end{equation}
We use Eqs.(\ref{6}) and (\ref{9}) in Eq.(\ref{21}), and get the following expression
\begin{equation}\label{22}
1+\omega_{1}(z)=\frac{1}{\frac{9}{2}(1-4\kappa\lambda)\bigg(\frac{\beta_{1}+(c^2(0)-\beta_{1})(1+z)^2}
{\frac{\beta_{1}q(z)}{(1+z)^2}
(\beta_{1}-\beta_{1}(1+z)^{2}+(1+z)^{2}c^{2}(0))}\bigg)+3\kappa\lambda},
\end{equation}
here $q(z)$ denotes the deceleration parameter. After evaluating it at present time, one can find an expression
for $\beta_{1}$, which is of the form
\begin{eqnarray}\nonumber
\beta_{1}&=&\bigg(1-4\kappa\lambda+\omega_{1}(0)(1-4\kappa\lambda)+q(0)(1
-3\kappa\lambda-3\kappa\lambda\omega_{1}(0)+\frac{2}{9})\bigg)\\\label{23} & \times & \bigg(\frac{c^2(0)}{(-\frac{2}{9}+3\kappa\lambda+3\kappa\lambda\omega_{1}(0))}\bigg).
\end{eqnarray}
Taking $q(0)\thicksim-0.5$ and $\beta_{1}\thickapprox\frac{3}{2}c^2(0)$, we obtain the following equivalence
\begin{equation}\nonumber
\omega_{1}(0)\thickapprox\bigg(\frac{11-63\kappa\lambda}{-9+63\kappa\lambda}\bigg).
\end{equation}
If we take large value of $\lambda$ then result approach to cosmological constant $\omega_{1}(0)=-1$ and for
small value of $\lambda$, the model lies in phantom era. Using Eq.(\ref{21}), the
expression of $q(z)$ at present time can be constrained as under
\begin{equation}\label{24}
q(0)\lesssim\frac{1}{(3\kappa\lambda-3\kappa\lambda\omega_{1}(0)+\frac{2}{9})}\bigg(-\frac{4}{3}+\frac{17}{2}
\kappa\lambda+\omega_{1}(0)(\frac{17}{2}\kappa\lambda-1)\bigg),
\end{equation}
which verifies the accelerating expansion of the cosmos as remained in negative region for all the
specified values of the model parameters.

In the next section, we will examine the interacting case mathematically and graphically.

\section{Dynamics of Interacting Model}

In this section, we describe the cosmological solutions for two components that are not conserved
separately but there exists an interaction between them. For curved FLRW space-time, $H_{n}^{2}(z)$
can be written as
\begin{equation}\label{25}
H_{n}^{2}(z)=\frac{1}{3H_{0}^2}(\rho_{1}+\rho_{2})\bigg(\frac{\kappa-3\kappa^2\lambda(1+\omega)}
{(1-4\kappa\lambda)(1-\Omega_{K}(z))}\bigg).
\end{equation}
The corresponding conservation equations in the form of $\rho_{1}$ are as follows
\begin{eqnarray}\label{26}
\rho_{1}^\prime\bigg(\frac{1-3\kappa\lambda-3\kappa\lambda\omega_{1}}{1-4\kappa\lambda}\bigg)-3
\bigg(\frac{1+\omega_{1}}{1+z}\bigg)\rho_{1}&=&-Q,\\\label{27} (r\rho_{1})^\prime\bigg(\frac{1-3\kappa\lambda-3\kappa\lambda\omega_{2}}{1-4\kappa\lambda}\bigg)
-3\bigg(\frac{1+\omega_{2}}{1+z}\bigg)r\rho_{1}&=&Q,
\end{eqnarray}
here $Q$ be the interaction term, which is assumed to be
\begin{equation}\nonumber
Q(z)=b^2H(z)\rho_{1}.
\end{equation}
Since $\rho_{1}=3c^{2}(z)H^{2}(z)$. Putting expression of $Q(z)$ in Eq.(\ref{27}), and taking $\omega_{2}=0$,
we are able to evaluate $\rho_{2}(z)$ given as follows
\begin{eqnarray}\nonumber
\rho_{2}(z)&=&2^{F_{1}}\bigg(3,\frac{3(C-(\frac{1-4\kappa\lambda}{1-3\kappa\lambda}))+1}{C+1};\frac{3(C-(\frac{1-4\kappa\lambda}{
1-3\kappa\lambda}))+C+2}{C+1};\frac{B(z+1)^{C+1}}{D}\bigg)\\\label{28}&\times&
\bigg(\frac{3b^2c^2(C+1)^3}{(1+z)^{-3(\frac{1-4\kappa\lambda}{1-3\kappa\lambda})}}\bigg)\bigg(\frac{1-4\kappa\lambda}
{1-3\kappa\lambda}\bigg)\bigg(\frac{(z+1)^{3(C-(\frac{1-4\kappa\lambda}{1-3\kappa\lambda}))+1}}{3(C-(\frac{1-4\kappa\lambda
}{1-3\kappa\lambda})+1)D^3}\bigg),
\end{eqnarray}
where
\begin{equation}\nonumber
A=\frac{1-4\kappa\lambda}{1-3\kappa\lambda(1+\omega_{1})},\quad B=\frac{Ab^2}{2},\quad C=\frac{3}{2}(1+\omega_{1})A,
\quad D=\big(1-\frac{B}{C+1}\big)(C+1).
\end{equation}
For interacting case, we treat $c^{2}$ as a constant term in order to evaluate the exact solutions of the model.

\begin{figure}\centering
\epsfig{file=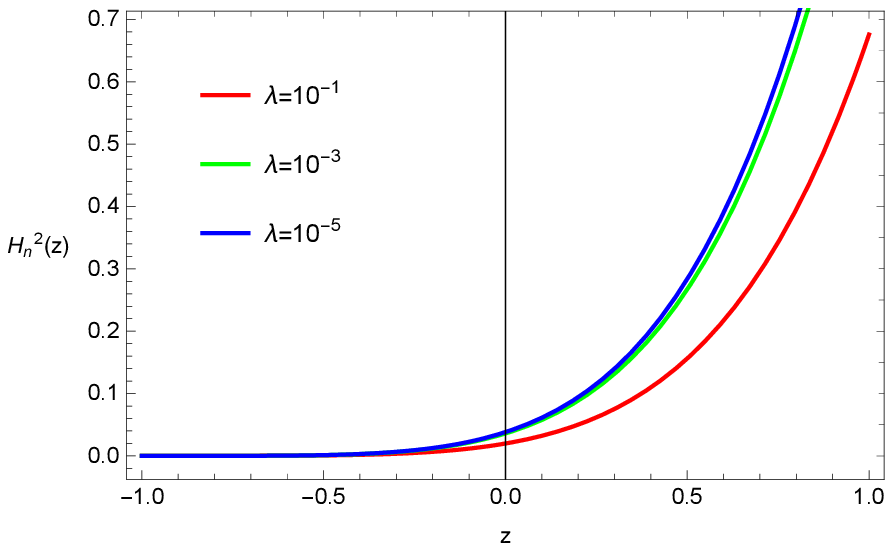, width=0.55\linewidth}\epsfig{file=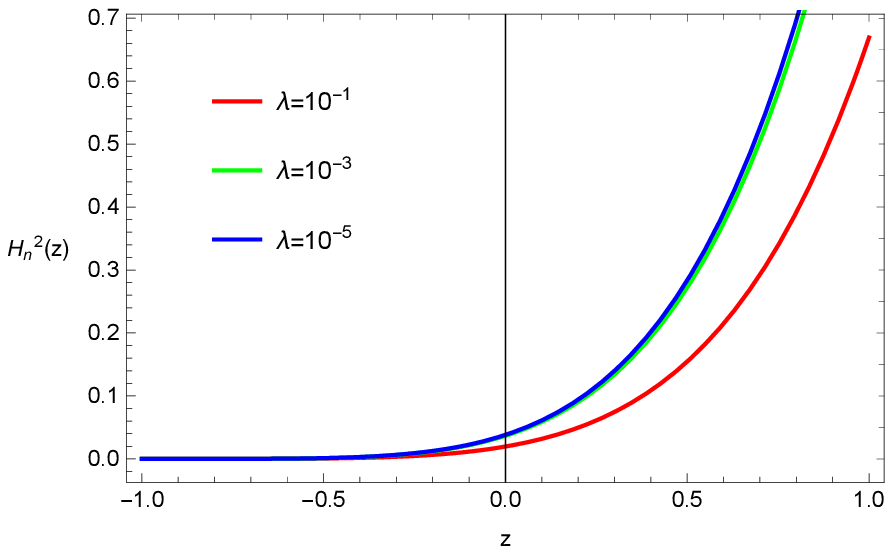, width=0.55\linewidth}
\caption{Graph of $H_{n}^{2}(z)$ versus $z$. Left plot for $\Omega_{K}(0)=+0.005$ and right plot
for $\Omega_{K}(0)=-0.005$. Values of other parameters are $c^{2}(0)=0.63,~\kappa=1,~\omega=-1.03,~
\omega_{1}=-0.666486,~b=0.2,~H_{0}=67.4$.  \label{f4}}
\end{figure}

\begin{figure}\centering
\epsfig{file=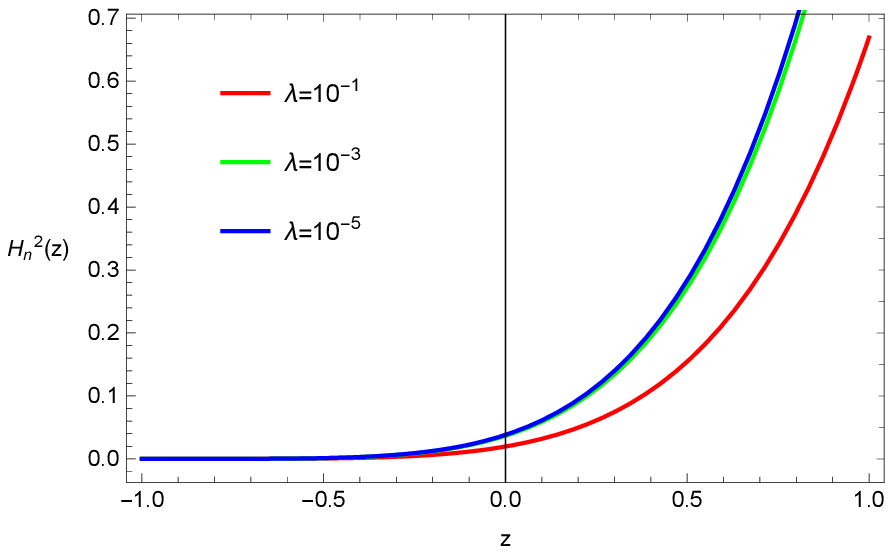, width=0.55\linewidth}\epsfig{file=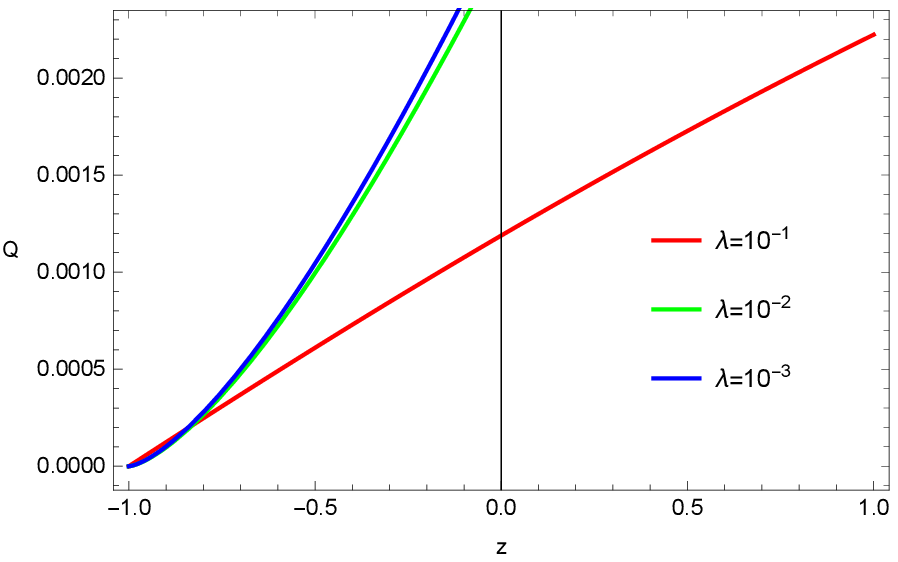, width=0.55\linewidth}
\caption{Left graph of $H_{n}^{2}(z)$ versus $z$ for $\Omega_{K}(0)=0$. Right graph of $Q$ versus $z$ \label{f5}}
\end{figure}

Putting these values of $\rho_{1}$ and $\rho_{2}$ in Eq.(\ref{25}), we obtain exact solution for $H_{n}^{2}(z)$
in the following form
\begin{eqnarray}\nonumber
H_{n}^{2}(z)&=&\frac{1}{3H_{0}^2}\bigg(\frac{\kappa-3\kappa^2\lambda-3\kappa^2\lambda\omega}{(1-4\kappa\lambda)(1-\Omega_{K}(z))}\bigg)
\bigg(3c^2H^2(z)+\bigg(\frac{3b^2c^2(C+1)^3}{(1+z)^{-3(\frac{1-4\kappa\lambda}{1-3\kappa\lambda})}}\bigg)\\\nonumber&\times&  2^{F_{1}}\bigg(3,\frac{3(C-(\frac{1-4\kappa\lambda}{1-3\kappa\lambda}))+1}{C+1};\frac{3(C-(\frac{1-4\kappa\lambda}{
1-3\kappa\lambda}))+C+2}{C+1};\frac{B(z+1)^{C+1}}{D}\bigg)\\\label{29}&\times&
\bigg(\frac{1-4\kappa\lambda}
{1-3\kappa\lambda}\bigg)\bigg(\frac{(z+1)^{3(C-(\frac{1-4\kappa\lambda}{1-3\kappa\lambda}))+1}}{3(C-(\frac{1-4\kappa\lambda
}{1-3\kappa\lambda})+1)D^3}\bigg)\bigg).
\end{eqnarray}
Now we describe the behavior of the Eq.(\ref{29}) graphically by varying $\lambda$ for $\Omega_{K}(0)=\pm0.005$ and
$\Omega_{K}(0)=0$. We use values of the model parameters as described in the previous section. In this equation,
$b^{2}$ be the coupling term that must satisfy the constraint, $|b^{2}|<1$. The other parameter is $\omega_{1}$
whose value at the present time must be less than -0.666485 or -0.666007 as given in \cite{24}. We can see from
Figs. \textbf{\ref{f4}} and \textbf{\ref{f5}} (left plot) that as $z$ evolves from past to future era, the trajectories of
$H_{n}^{2}(z)$ positively increasing for open, closed and flat universe, respectively. For $\lambda=0$, graph
shows similar behavior for open, closed and flat universe.
In order to check the behavior of $Q(z)$, we evaluate its expression by putting $\rho_1(z)$ and $H_{n}^{2}(z)$. The graphical
behavior of this term is presented in Fig. \textbf{5} (right plot). It can be seen that the obtained interaction term goes
on increasing from past to future era of the evolution.

Now we calculate the mathematical expression of $r(z)$ for interacting model as
\begin{equation}\label{30}
r(z)=\frac{\rho_{2}(z)}{3c^2H^{2}_{n}(z)H_{0}^2}.
\end{equation}
\begin{figure}\centering
\epsfig{file=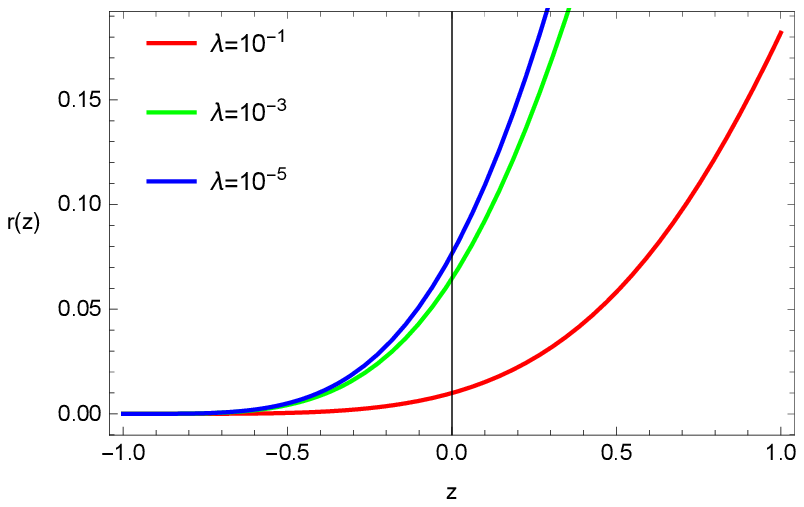, width=0.550\linewidth}\epsfig{file=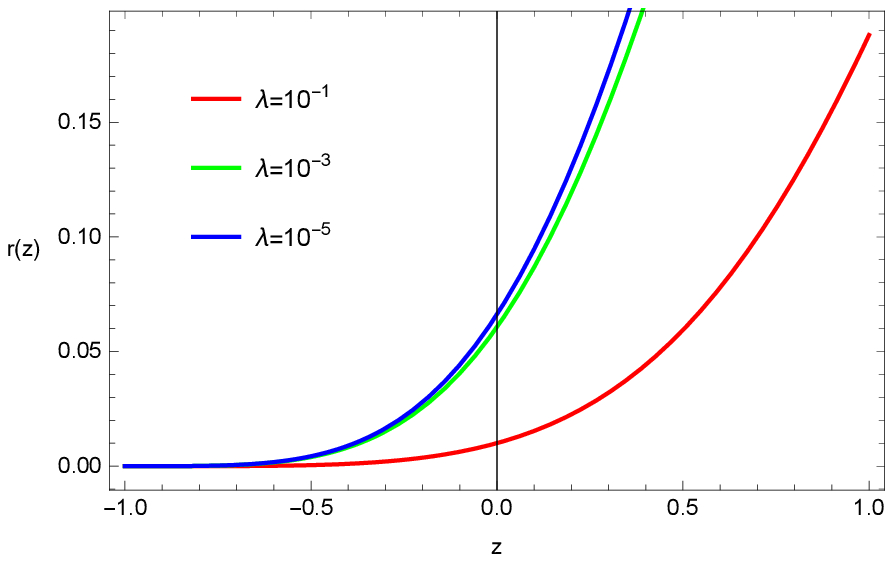, width=0.550\linewidth}
\caption{Graph of $r(z)$ versus $z$. Left plot for $\Omega_{K}(0)=+0.005$ and right plot for $\Omega_{K}(0)=-0.005$. \label{f6}}
\end{figure}
\begin{figure}\center
\epsfig{file=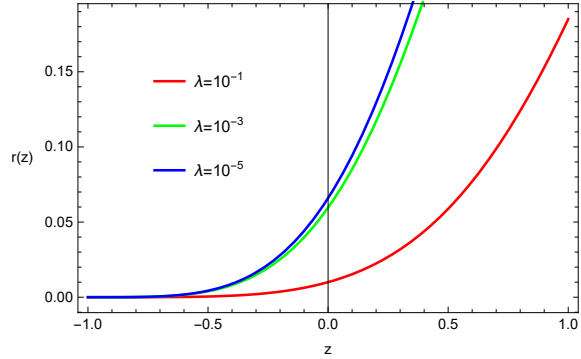, width=0.550\linewidth}
\caption{Graph of $r(z)$ versus $z$. For $\Omega_{K}(0)=0$. \label{f7}}
\end{figure}
We can obtain the final result of $r(z)$ by using the solutions of $\rho_{2}(z)$ and
$H_{n}^{2}(z)$ given in Eqs.(\ref{28}) and (\ref{29}), respectively in Eq.(\ref{30}).
The graphical illustration of $r(z)$ versus $z$ is given in Figs. \textbf{\ref{f6}}
and \textbf{\ref{f7}} for open, closed and flat universe, respectively. This parameter
for different values of $\lambda$ exhibits the same physical behavior as we have discussed
earlier in non-interacting model.

The EoS parameter $\omega_{1}(z)$, for interacting model has the following form
\begin{equation}\label{31}
1+\omega_{1}(z)=\frac{-1-q(0)+b^2H_{0}(1-4\kappa\lambda)}{-3\kappa\lambda(4+q(0)-12\kappa\lambda)}.
\end{equation}
It has been checked that by fixing $H_0,~q(0)$, $\omega_{1}(z)$ approaches to cosmological constant model for
large values of $\lambda$ whereas for small $\lambda$, the considered model behaves like phantom model.

\section{Cosmological Parameters for Non-interacting and Interacting Models}

In this section, we constraint the involved model parameters to attain $\Lambda$CDM limit via evaluated
cosmological parameters like deceleration parameter $q(z)$, statefinder pair $(j,s)$ and $Om(z)$-diagnostic.

\subsection{Periodicity of $q(z)$}

The parameter $q(z)$ is one of the parameter that can describe the dynamical nature of the universe in
geometrical sense. It distinguishes between two phases of the universe, $-1\leq q<0$ shows accelerated
phase and $q\geq 0$ leads to decelerated phase of the universe. Exact time dependence of this parameter
still not found. Some specific values of $q$ represent specific eras like de-Sitter model is attained for
$q=-1$, power-law model for $-1<q<0$ and super-exponential model can be achieved for $q<-1$. Mathematical
expression for $q(z)$ is as follows
\begin{equation}\label{32}
q(z)=-1-\frac{\dot{H}}{H^2},
\end{equation}
to find the value of $H^{2}(z)$, we consider Eq.(\ref{1}) as
\begin{equation}\nonumber
H^2(z)=\frac{1}{3(1-4\kappa\lambda)(1-\Omega_{K}(z)}\bigg(\rho_{1}\kappa(1+r(z))(1-3\kappa\lambda(1+\omega))\bigg).
\end{equation}
Since $\dot{H}=-\frac{\kappa}{2}(\rho+P)+\frac{K}{a^{2}}$, after simplifications, we get the following expression
for $\dot{H}(z)$
\begin{equation}\nonumber
\dot{H}=-\frac{\kappa}{2}\rho_{1}(1+r(z))+\frac{K(1+z)^2}{a_{0}^2}.
\end{equation}
Using above two expressions in Eq.(\ref{32}), we get the expression of $q(z)$ for non-interacting case as
\begin{eqnarray}\nonumber
q(z)&=&-1+\frac{(1+z)^3\kappa}{a_{0}^2}-\frac{3(-1+\Omega_{K}(0)(1+z)^2)}{2(c^2(0)(1+z)^2+\beta_{1}z(2+z))}
\\\label{33}&\times&\frac{(3c^2(0)(1+z)^2+\beta_{1}z(2+z))(-1+4\kappa\lambda)}{(1-3\kappa\lambda)(1+\omega)}.
\end{eqnarray}
\begin{figure}
\epsfig{file=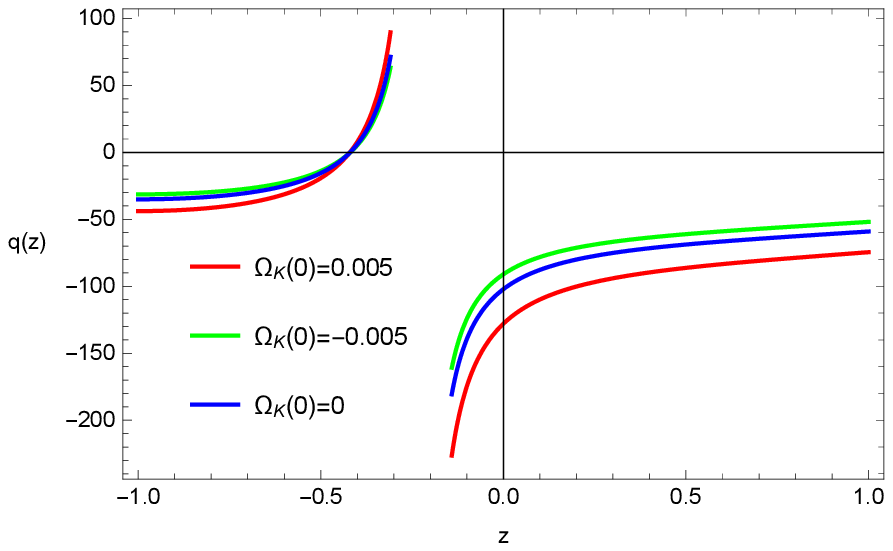, width=0.50\linewidth}\epsfig{file=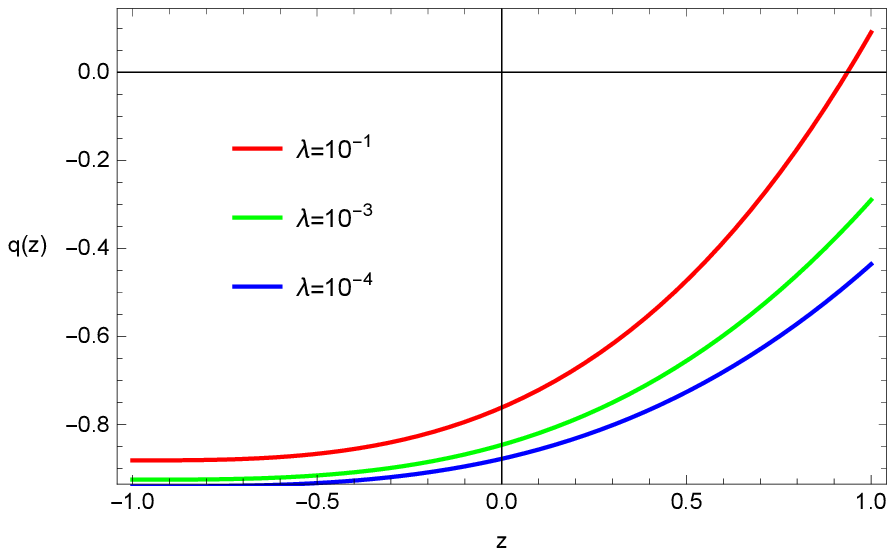, width=0.50\linewidth}
\caption{Graph of $q(z)$ versus $z$. Left plot for non-interacting model by fixing $\lambda=10^{-1}$ and
varying $\Omega_{K}(0)$; right plot for interacting model varying $\lambda$ and fixing $\Omega_{K}(0)=0.005$. \label{f8}}
\end{figure}
For interacting model, $q(z)$ has the following expression
\begin{equation}\label{34}
q(z)=-1+\frac{\frac{\kappa}{2}3c^2H^2(z)(1+r(z))+\frac{K(1+z)^2}{a_{0}^2}}{H^2(z)}.
\end{equation}
Taking the values of $H^{2}(z)$ and $r(z)$ from interacting model and putting them in Eq.(\ref{34}), we can obtain
the exact solution of $q$ in term of $z$ only. From left panel of Fig. \textbf{\ref{f8}}, we observed that in future era,
$q(z)$ is transiting from accelerating phase to decelerating phase of the universe for $\lambda=10^{-1}$. Whereas in the
past, it always lies in the accelerating phase for open, closed and flat universe. The remaining values are used from the
Tables \textbf{1} and \textbf{2} for non-interacting case. We plot $q(z)$ for interacting model in the right plot of
Fig. \textbf{\ref{f8}}. In this scenario, $q(z)$ shows transition from accelerating to decelerating phase for
$\lambda=10^{-1}$ only but for $\lambda=10^{-2},~10^{-3}$, it shows accelerating phase of the universe throughout for
all values of $\lambda$. In this case, we have checked that $q(z)$ behaves the same for open, closed and flat models
so we have added the graph for $\Omega_{K}(0)=0.005$ only.

\subsection{Periodicity of Statefinder Pair $(j,s)$}

First, we calculate the expression of $j$ for non-interacting model by given formula \cite{31}
\begin{equation}\label{33}
j=\frac{\ddot{H}}{H^3}-3q(z)-2,
\end{equation}
here the expression for $\ddot{H}$ is as follows
\begin{equation}\nonumber
\ddot{H}=-\frac{\kappa}{2}(\dot{\rho_{1}}+\dot{\rho_{2}})+\frac{2K}{a_{0}^2}(1+z)\dot{z}.
\end{equation}
With the help of $\ddot{H},~H^{3}$ and $q(z)$ that are evaluated for non-interacting model, we
can find the expression of $j$. For interacting model, $j$ has the following form
\begin{eqnarray}\nonumber
j&=&\bigg((1+z)\bigg(\frac{\kappa}{2}(\frac{1-4\kappa\lambda}{1-3\kappa\lambda})
(-bH(z)+\frac{3(1+\omega_{1})}{1+z})3c^2H^2(z)\bigg)
\\\nonumber&+&\bigg(\frac{\kappa}{2}(\frac{1-4\kappa\lambda}{1-3\kappa\lambda})
(3b^2c^2H^3(z)+\frac{3\rho_{2}}{1+z})-\frac{2K}{a_{0}^2}(1+z)\bigg)\bigg)(H^2(z))^{-1}
\\\label{37}&-&3q(z)-2.
\end{eqnarray}
On substituting the values of $q(z),~\rho_{2}(z),~H^{2}(z)$ and $H^{3}(z)$ from interacting model,
one can easily obtain the corresponding expression for $j$.

The snap parameter $s$ is calculated via following formula
\begin{equation}\label{34}
s=\frac{j-1}{3(q-\frac{1}{2})}.
\end{equation}
Using the value of $j$ and $q$ for non-interacting and interacting models, we can find the expression for
corresponding $s$, respectively. This order pair, $(j,s)=(1,0)$ shows $\Lambda$CDM model and $(j,s)=(1,1)$
shows CDM limit. Also, $j<1$ and $s>0$ represents the \textit{phantom} and \textit{quintessence} regions
for DE while $j>1$ and $s<0$ shows chaplygin gas behavior \cite{32}.
\begin{figure}
\epsfig{file=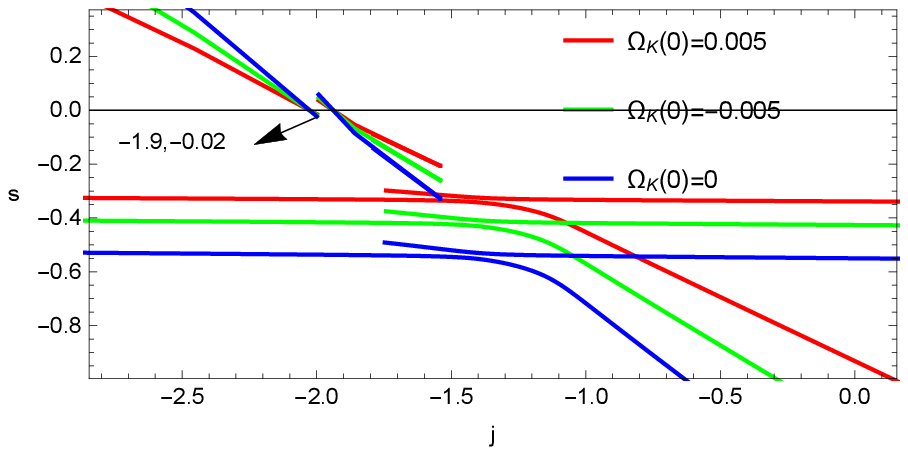, width=0.50\linewidth}\epsfig{file=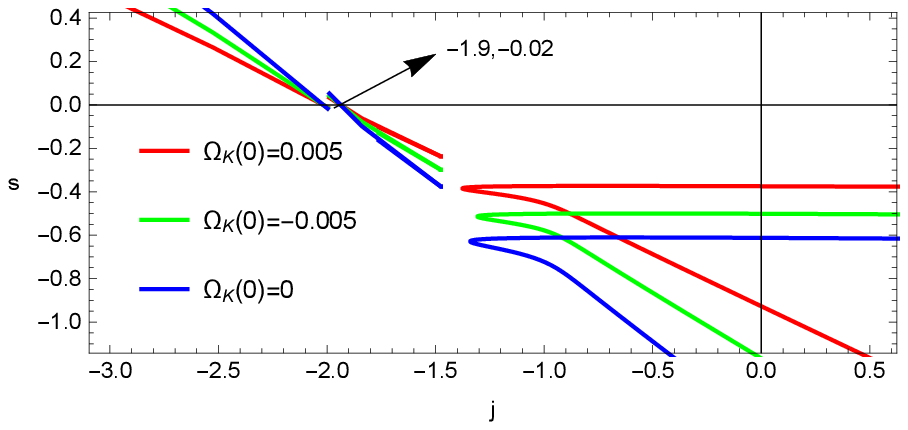, width=0.50\linewidth}
\caption{Parametric plot of $j-s$ for non-interacting model. Left plot for $\lambda=10^{-1}$ and right plot
for $\lambda=10^{-2}$. \label{f9}}
\end{figure}
Left panel of Fig. \textbf{\ref{f9}} is plotted for three different values of $\Omega_{K}(0)$ by fixing $\lambda=10^{-1}$.
It shows that this value of $\lambda$ did not generate any physical result because $j<1$ and $s<0$ is not matched any of
the regions that described above, so for non-interacting model, we are unable to guess in which region it will lie. Figure
\textbf{\ref{f9}} (right plot) produce the same results. So, this model is far from $\Lambda$CDM limit.
\begin{figure}
\epsfig{file=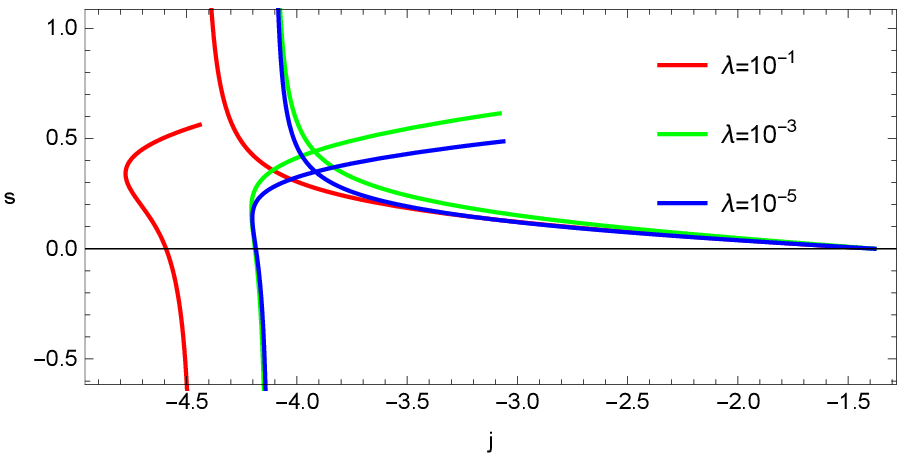, width=0.50\linewidth}\epsfig{file=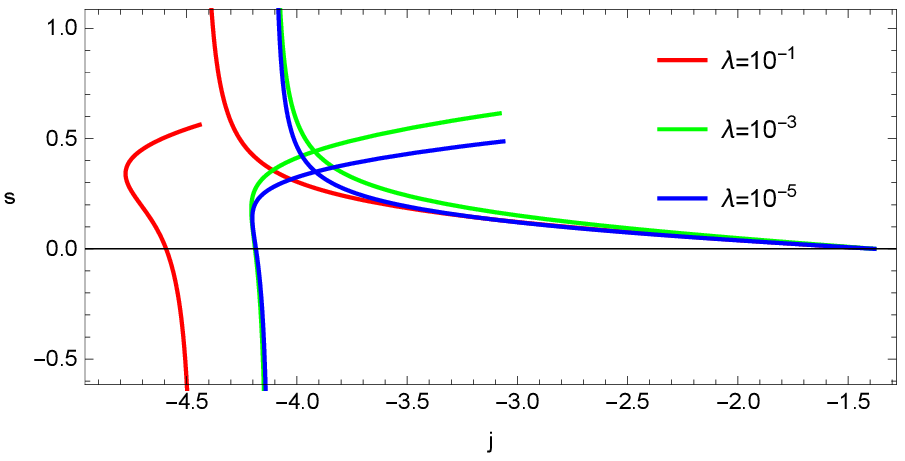, width=0.50\linewidth}
\caption{Parametric plot of $j-s$ for interacting model. Left plot for $\Omega_{K}(0)=0.005$ and right plot for
$\Omega_{K}(0)=-0.005$. \label{f10}}
\end{figure}
In Fig. \textbf{\ref{f10}} (left and right plots), we observed that both graph shows DE eras, i.e., \textit{phantom} and
\textit{quintessence} as $s>0$ and $j<1$, and both of the plots approach to $\Lambda$CDM limit. During interacting model,
we choose a different range for $z$, i.e., $z=(-1.5,1.5)$ because $z=\pm1$ are the singular points.

\subsection{$Om(z)$-diagnostic Parametrization for Non-interacting and Interacting Models}

Now, we calculate the $Om(z)$ parameter using following formula
\begin{equation}\nonumber
Om(z)=\frac{h^2(z)-1}{(1+z)^3-1};\quad h(z)=\frac{H(z)}{H_{0}}
\end{equation}
hence $Om(z)$ turn out to be
\begin{equation}\label{39}
Om(z)=\frac{H^2(z)-H_{0}^2}{H_{0}^2((1+z)^3-1))}.
\end{equation}
\begin{figure}\centering
\epsfig{file=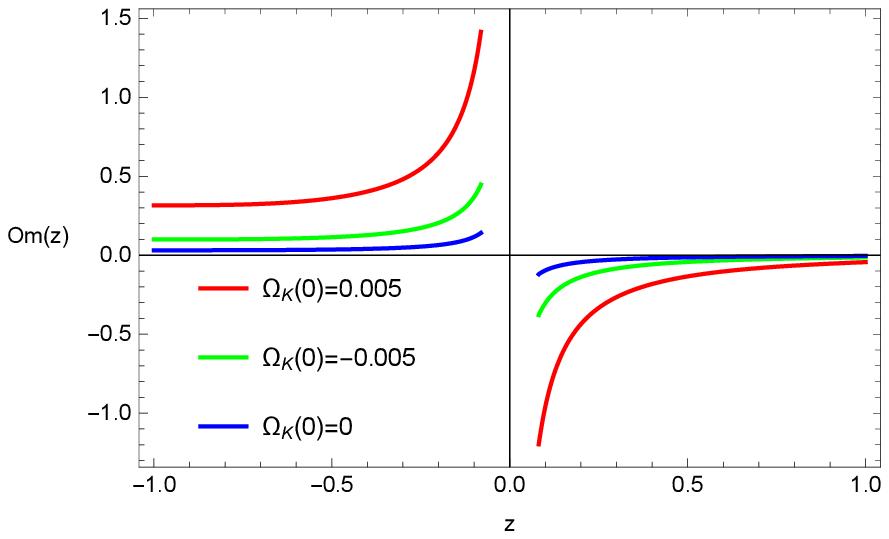, width=0.50\linewidth}\epsfig{file=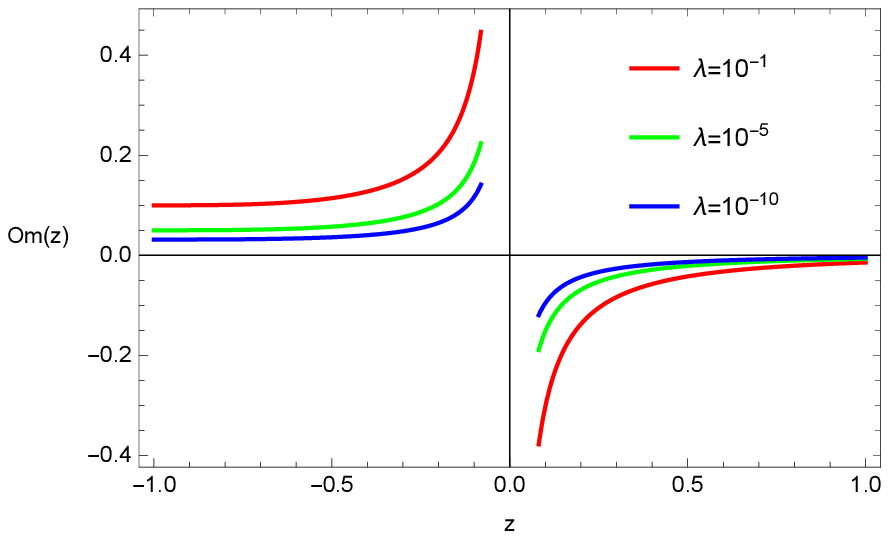, width=0.50\linewidth}
\caption{Plot of $Om(z)$-diagnostic versus $z$. \label{f11}}
\end{figure}

The final expressions of $Om(z)$ parameter in terms of $z$ only can be obtained by putting $H^{2}(z)$, which are 
calculated for both non-interacting and interacting models, alternatively. $Om(z)$-diagnostic is another effective 
geometric tool that can describe different phases of the universe. The universe is assumed to be passing through 
\textit{phantom} era for $Om(z)>0$ whereas negative region represents the \textit{quintessence} era of DE. In 
Fig. \textbf{\ref{f11}}, left panel is plotted for non-interacting model for three different values of $\Omega_{K}(0)$ 
keeping $\lambda=10^{-1}$ as fixed parameter, and right plot is for interacting model by varying $\lambda$. In the right 
plot, we take $\Omega_{K}(0)=+0.005$ for all the three trajectories. It is also checked that the other two values 
$\Omega_{K}(0)=-0.005,~0$, give the similar behavior. Both of the graph show positive slope in future epoch, which 
leads to \textit{phantom} era while negative slope in past epoch is showing the \textit{quintessence} era of the universe.

\section{Final Results}

The main motivation of this paper is to check the effectiveness of RG in describing the expanding nature of the universe 
at an accelerating rate. Such scenarios have been widely discussed in general relativity but there is a room to discuss it
in modified theories of gravity. Since RG is the most simplest generalization of GR because it reduced to GR when the 
additional parameter $\lambda$ is valued zero, therefore we decided to discuss cosmology in RG taking curved FLRW 
space-time. In this framework, the corresponding dynamical field equations and conservation equations have been 
formulated considering perfect fluid distribution. The HDE and DM are assumed to be components of the fluid, and 
further study has been distributed into two cases; non-interacting HDE-DM model and interacting HDE-DM model. Using 
HDE density as an IR cut-off, we have calculated the exact solutions of these dynamical equations involving, normalized
Hubble parameter ($H^{2}_n(z)$) and energy density of DM ($\rho_2(z)$), which further lead us to find out coincidence parameter 
$r(z)$. Using solution of $r(z)$ for non-interacting case, we have evaluated the holographic parameter $c^{2}(z)$, which must 
lie in a specific range $c^{2}(0)=[0.63,0.66]$ at present time. This range has been undertaken in the far future and also 
satisfied in the past. For interacting case, we have considered it a constant $c^2$ term in order to evaluate the exact solutions.

In both of the cases, the obtained solutions have been analyzed graphically. The detail is as follows:    
The expressions of $H_{n}^{2}(z)$ are strongly depended upon two main parameters $\lambda$ and $\Omega_{K}(0)$. The remaining 
model parameters are constraint in Table \textbf{1} and \textbf{2}.
\begin{itemize}
\item For non-interacting case, the result is plotted in Figs. \textbf{\ref{f1}} and \textbf{\ref{f2}} for open, closed and flat 
universe. It is observed that $H_{n}^{2}(z)$ remains positive and less than unity for all values of $0\leq\lambda<1$ except left 
plot of Fig. \textbf{\ref{f1}} for $\lambda=10^{-1}$, where $H_{n}^{2}(z)<0$ in the future era.
\item $H_{n}^{2}(z)$ for interacting case has been plotted in Figs. \textbf{\ref{f4}} and \textbf{\ref{f5}} (left plot), which exhibit 
physical behavior from past epoch to future epoch for closed, open and flat universe.
\end{itemize}
The expression of $r(z)$ depends upon $c^{2}(0),~\Omega_{K}(0),~\lambda$ and $\omega(0)$. 
\begin{itemize}
\item Its behavior for non-interacting and interacting models has been shown in Figs. \textbf{\ref{f3}} and \textbf{\ref{f6}}, 
\textbf{\ref{f7}}, respectively. From all the graphs, it is clear that $r(z)$ evolutes from past to future era taking various 
values of $\lambda$ and lies below unity for open, closed and flat universe models. This resolves the cosmic coincidence issue.
For non-interacting case, we also evaluated the exact solution of the interaction term $Q(z)$, and checked that it obeys the 
physical circumstances (see right plot of Fig. \textbf{\ref{f5}}).  
\end{itemize}

Further, we have calculated essential cosmological parameters that can describe the nature of the universe more efficiently.
The graphical description is as follows:
\begin{itemize}
\item The declaration parameter has been plotted for non-interacting and interacting models in left and right plots of 
Fig. \textbf{\ref{f8}}, respectively. From left graph, it is observed that $q(z)$ moves from accelerating phase to 
decelerating phase in the future era for $\Omega_{K}(0)=\pm0.005,~0$ taking $\lambda=10^{-1}$. For other values of $\lambda$,
$q(z)<0$. While right panel depicts that $q(z)$ always stays in the accelerating phase of the universe for all values of 
$\lambda\ll1$ and $\Omega_{K}(0)=\pm0.005,~0$.
\item For non-interacting model (Fig. \textbf{\ref{f9}}), we are unable to describe the behavior of the universe because 
in this case, we attain the specific region where $j<1$ and $s<0$, which did not match to any physical region of the universe. 
So this parametric plot $j-s$ is not physical for all values of $\lambda$.
\item For interacting model (Fig. \textbf{\ref{f10}}), we obtained a physical region $s>0$ and $j<1$ that describe the 
DE eras (\textit{phantom} and \textit{quintessence}). The $j-s$ trajectories for interacting model also approach to 
$\Lambda$CDM limit. So, we can say that an interacting model is most physical than non-interacting one in this respect.
\item For both non-interacting and interacting models $Om(z)$-diagnostic depicts both \textit{phantom} and \textit{quintessence} 
epoches of the universe (see Fig. \textbf{\ref{f11}}).
\end{itemize}
It is worthy to mention here that our results are in perfect agreement with GR taking Rastall parameter $\lambda=0$ as discussed 
in \cite{30} only for non-interacting model. We would like to mention here that the results are also compatible with \cite{24}, 
where we have discussed this idea of working in modified theory of gravity as a future work.


\begin{thebibliography}{40}

\bibitem{1} T. Clifton, P.G. Ferreira, A. Padilla and C. Skordis, Phys. Rep. \textbf{513}(2012)1-189.
\bibitem{2} S. Permutter, et al., Astrophys. J. \textbf{517}(1999)565-598.
\bibitem{3} M. Tegmark, et al., Phys. Rev. D \textbf{69}(2004)103501-103520.
\bibitem{4} D.N. Spergel, et al., Astrophys. J. Suppl. Ser. \textbf{148}(2003)175-225.
\bibitem{5} N. Aghanim, et al., Astro. Astrophys. \textbf{641}(2020)A6.
\bibitem{6} P.J.E. Peebles and B. Ratra, Rev. Mod. Phys. \textbf{75}(2003)559-606.
\bibitem{7} R.R. Caldwell, Phys. Lett. B \textbf{545}(2002)23-29.
\bibitem{8} Y. Aditya, S. Mandal, P.K. Sahoo and D.R.K. Reddy, Eur. Phys. J. C \textbf{79}(2019)1020-1036.
\bibitem{9} C.A. Picon, V. Mukhanov and P.J. Steinhardt, Phy. Rev. D \textbf{63}(2001)103510-103522.
\bibitem{10} H. kim, H.W. Lee and Y.S. Myung, Phys. lett. B \textbf{632}(2006)605-609.
\bibitem{11} A.K. Yadav, F. Rahman, S. Ray and G.K. Goswami, Eur. Phys. J. C \textbf{10}(2012)127-140.
\bibitem{12} L. Susskind, J. Math. Phys. \textbf{11}(1995)6377-6396.
\bibitem{13} M.H. Belkacemi, Z. Bouabdallaoui, M.B. Lopez, A. Errahmani and T. Ouali, Int. J. Mod. Phys. D \textbf{9}(2020)2050066-2050080.
\bibitem{28} A.G. Cohen, D.B. Kaplan and A.E. Nelson, Phys. Rev. Lett. \textbf{82}(1999)4971-4974.
\bibitem{15} M. Li, Phy. Lett. B \textbf{603}(2004)1-5.
\bibitem{16} M. Sharif, A. Shah and K. Bamba, Sym. \textbf{10}(2018)153-163.
\bibitem{14} A.A. Mamon, A.H. Ziaie and K. Bamba, Mod. Phys. Lett. A \textbf{30}(2020)2050251-2050257.
\bibitem{17} W.A.G. De Moraes and A.F. Santos, Gen. Rel. Grav. \textbf{51}(2019)167-184.
\bibitem{18} P. Rastall, Phys. Rev. D \textbf{6}(1972)3357-3359.
\bibitem{19} P. Rastall, Can. J. Phys. \textbf{54}(1976)66-75.
\bibitem{20} D. Das, S. Dutta and S. Chakraborty, Eur. Phys. J. C \textbf{78}(2018)810-818.
\bibitem{21} S. Ghaffari, A.A Mamon, H. Moradpour and A.H. Ziaie, Mod. Phys. Lett. A \textbf{35}(2020)2050276-2050290.
\bibitem{22} E. Karimkhani and A.K. Mohammadi, Astrophys. Space Sci. \textbf{364}(2019)177-209.
\bibitem{23} F. Arevalo, P. Cifuentes, S. Lepe and F. Pena, Astrophys. Space Sci. \textbf{2}(2014)899-907.
\bibitem{24} R. Saleem and M.J. Imtaiz, Class. Quan. Gravit. \textbf{37}(2020)065018-065036.
\bibitem{25} S.D. Campo, J.C. Fabris, R. Herrera and W. Zimdhal, Phys. Rev. D \textbf{83}(2011)123006-123020.
\bibitem{26} A.A. Costa, R.C.G. Landim, B. Wang and E. Abdalla, Eur. Phys. J. C \textbf{78}(2018)746-751.
\bibitem{27} E. Aydiner, Sci. Rep. \textbf{1}(2018)1-12.
\bibitem{29} M. Li, Phys. Lett. B \textbf{603}(2004)1-5.
\bibitem{30} M. Cruz and S. Lepe, Eur. Phys. J. C \textbf{78}(2018)994-999.
\bibitem{31} V. Sahni, T.D. Saini, A.A. Starobinsky and U. Alam, J. Exp. Theor. Phys. Lett. \textbf{77}(2003)201-206.
\bibitem{32} A. Jawad, S. Rani, S. Saleem, K. Bamba and R. Jabeen, Sym. \textbf{11}(2019)1009-1026.
\bibitem{33} M. Shahalam, S. Sami and A. Agarwal, Mon. Not. Roy. Astron. Soc. \textbf{448}(2015)2948-2959.
\bibitem{34} D. Bak and S.J. Ray, Class. Quan. Grav. \textbf{15}(2002)83-92.
\bibitem{35} N. Radicella and D. Pavon, J. Cosmol. Astropar. Phys. \textbf{10}(2010)005-012.
\bibitem{36} H.E.S. Velten, R.F. Vom Marttens and W. Zimdahl, Eur. Phys. J. C \textbf{11}(2014)1-8.
\end{thebibliography}
\end{document}